\documentclass[superscriptaddress,aps,prd, 10pt,twocolumn]{revtex4-1}
\usepackage{epsfig}
\usepackage{graphicx}
\usepackage{amsmath}
\usepackage[latin1]{inputenc}
\usepackage{amsbsy}
\usepackage{color}
\usepackage{epsfig}
\usepackage{graphicx}
\usepackage{amsmath}
\usepackage{amssymb}
\usepackage{bbm}
\usepackage{amsbsy}
\usepackage{verbatim}
\usepackage{slashed}\usepackage{natbib}
\newcommand{\ee}{\end{equation}}
\newcommand{\bb}{\begin{equation}}
\newcommand{\eqb}{\begin{eqnarray}}
\newcommand{\eqf}{\end{eqnarray}}

\newcommand{\1}{{\'{\i}}}

\def\1{\'{\i}}
\newcommand{\tr}{\operatorname{tr}}

\usepackage{ulem}

\begin{document}
\title{Vacuum instability by a chromoelectric field in $2+1$-dimensions}

\author{M. Loewe}
\affiliation{Departamento    de   F\'{\i}sica,    Pontificia   Universidad
  Cat\'olica de Chile, Casilla 306, Santiago 22, Chile}
\affiliation{Centre for Theoretical Physics and Mathematical Physics, University of Cape Town, Rondebosch 7700, South Africa}
\author{F. Marquez}
\affiliation{Departamento    de   F\'{\i}sica,    Pontificia   Universidad
  Cat\'olica de Chile, Casilla 306, Santiago 22, Chile}

\author{R. Zamora}
\affiliation{Departamento    de   F\'{\i}sica,    Pontificia   Universidad
  Cat\'olica de Chile, Casilla 306, Santiago 22, Chile}

\begin{abstract}
The problem of vacuum instability and creation of pairs by external fields $SU(2)$ is studied. The effective mass operator has complex eigenvalues implying decay and pair creation. We consider a constant external chromoelectric field a find a region in the parameter space where the vaccum becomes unstable and decays in pairs. Although the calculation is done in $SU(2)$ color group it is also valid for $SU(3)$.
\end{abstract}
%\pacs{PACS numbers:}
%\date{\today}
\maketitle

\section{Introduction}

The prediction of pair creation by a constant background electric field is an important nonperturbative result in quantum electrodynamics (QED). It shows that the vacuum of the quantum field theory (QED) acquires a nontrivial and nonperturbative unstable character in such backgrounds \cite{schwinger, monton01, monton02, monton03,monton04} and, consequently, decays through pair creation. This instability manifests in the effective action of the theory having an imaginary part. Although the magnitude of  this effect is very small for field strengths that can be produced in the laboratory presently, it is expected that in the coming years very strong electric fields may be produced in the laboratory. This would allow for a direct verification of Schwinger's prediction. 

Vacuum instability and pair creation in an external field is also an interesting topic in itself to study in non-Abelian theories such as quantum chromodynamics (QCD). However, there are some important differences between an Abelian and a non-Abelian gauge theory. For example, the field strength tensor (in a matrix form) in a non-Abelian theory is defined as
\begin{equation}
F_{\mu\nu} = \partial_{\mu}A_{\nu} - \partial_{\nu}A_{\mu} + ig [A_{\mu}, A_{\nu}],\label{fs}
\end{equation}
where $A_{\mu}$ denotes the vector potential and $g$ the coupling constant so that it is not merely the curl of the vector potential. Under a non-Abelian gauge transformation
\begin{equation}
A_{\mu} \rightarrow U(x) A_{\mu} U^{-1} (x) - \frac{1}{ig} \left(\partial_{\mu} U(x)\right) U^{-1} (x),
\end{equation}
the field strength tensor is not invariant, rather it transforms covariantly as
\begin{equation}
F_{\mu\nu} \rightarrow U (x) F_{\mu\nu} U^{-1} (x).
\end{equation}
Since the field strength tensor is not gauge invariant (unlike in QED), it is not an observable and an external chromoelectric or chromomagnetic field is not a meaningful concept. It is also well known that in non Abelian gauge theories there exists field tensors which are realized in terms of different gauge field configurations which are not gauge equivalent. Also, Wilson loops calculated using these non equivalent gauge fields, produce different results \cite{Dietrich01}. Nonetheless it is an interesting question to study since, as in any collective phenomenon, such instabilities may arise  in the quark-gluon phase \cite{varios2}. Within the context of effective actions, we note that the effective action of the theory is a gauge invariant quantity since the fermion determinant $\det (i\slashed{D} -m)$, which leads to the effective action, is gauge invariant.

In this sense one can study the problem of a (color) charged particle moving in the presence of an external chromoelectric or a chromomagnetic field. Furthermore, as in the case of QED, one can assume the external fields to be constant \cite{Gavrilov01}. It was shown many years ago by Brown and Weisberger \cite{brown,singleton}  that  a constant chromoelectric field (or chromomagnetic) can be obtained from  two different classes of gauge potentials, namely, a) $A \propto x$ or , b) $A$ is  constant. Basically, the two classes correspond respectively to either the commutator or the curl in \eqref{fs} vanishing. The first class of potentials is familiar from the study of Abelian theories where a constant field can be associated with a potential of the form (symmetric gauge) $A_{\mu} = - \frac{1}{2} F_{\mu\nu} x^{\nu}$. The second class genuinely arises in a non-Abelian theory. The first class of potentials where the potential is linear in the coordinates has been used by many authors \cite{Nayak01, Nayak02, Nayak03, Nayak04, Nayak05}. In particular Saviddy \cite{savi} has used this choice of the potential in discussing the vacuum instability in quantum chromodynamics and Nielsen and Olesen in \cite{nielsen} have employed an analogous gauge to study the instability of $SU(2)$ Yang-Mills theory coupled to Higgs.
 
In this note we analyze the problem of vacuum instability for the case of constant gauge field strengths belonging to the non-Abelian $SU(2)$ group in $2+1$ dimensions \cite{nair, Huang01, Dietrich02, Gies02}. This same kind of potentials have been used in exploring temperature and density effects in the phrame of a discussion of phase transitions \cite{Klimenko01, Klimenko02}. In particular, we calculate the probability density for pair creation by external chromoelectric fields. Our analysis is valid for the case of a chromoelectric field. In the case of an external chromomagnetic field the effective action does not develop poles and the effect will be only to produce a condensed state $\langle\bar{\psi}\psi\rangle$. So, in this case we will not expect to observe any instability in the vacuum \cite{Gies02}. We have chosen to analyze the $2+1$ dimensional problem because it has already been shown that there is no instability for the $3+1$ dimensional case \cite{brown}. Also, $2+1$ dimensional discussion can be of relevance if we are considering quark-gluon plasma at very high temperature, $T\rightarrow \infty$, when one of the dimensions is expected to compactify  \cite{Gross:1980br}.

\section{Probability density for pair emission.}

As we have already mentioned, a constant non-Abelian field strength can be described by gauge potentials that are either linearly proportional to $x^{\mu}$ or by constant non-commuting gauge potentials. Many people have already used the first class of gauge potentials in studying the instability phenomena. In our calculation, however, we choose the second class of constant potentials since it provides an alternative and physical way of looking at the problem of vacuum decay. Let $A_{\mu} = A_{\mu}^{a} \tau^{a}, a=1,2,3$ denote the $SU(2)$ gauge potentials where $a$ denotes the color index and $\tau^{a} = \lambda_{a}/2$ are the $SU(2)$ color generators (in the fundamental representation). We consider a constant chromoelectric field along $x$-direction (spatial) pointing along the $3$-direction in the color space, namely,
\begin{equation}
E_{1} = F_{01} = ig [A_{0}, A_{1}] = E_{1}^{3} \tau_{3}.
\end{equation}
A simple choice of constant gauge potentials that yields such a chromoelectric field is given by $A_{0} = - \sqrt{\epsilon_{1}} \tau_{1}, A_{1} = \sqrt{\epsilon_{2}} \tau_{2}$ where $\epsilon_{1}, \epsilon_{2}$ are (dimensional) constants, leading to
\begin{equation}
F_{01} = E_{1} = E_{1}^{3} \tau_{3} = g \sqrt{\epsilon_{1}\epsilon_{2}} \tau_{3} = \epsilon \tau_{3}.\label{fs1}
\end{equation}
We note here that in $2+1$ dimensions both $g$ and $A$ have the dimensions of square root of a mass. Also, in $2+1$ dimensions there are two possible mass terms for the Lagrangian \cite{Pisarski01}. However one of these terms violates parity and time reversal while the other one violates chiral symmetry. In this article we work whithin a parity conserving model in $2+1$ dimensions \cite{Gies01}, where we work in a reducible representation of the gamma matrices \cite{Raya01}.\\

We can now study the motion of a four component Dirac fermion (belonging to a reducible representation) in such a background described by the equation
\begin{equation}
\left(i\slashed{D} - m\right)\psi = \left(i\slashed{\partial} - g \slashed{A} - m\right) \psi = \left(i\slashed{\partial} - M\right) \psi= 0.
\end{equation}
Here $\slashed{A} = \gamma^{\mu} A_{\mu}$, the three gamma matrices are defined in terms of the Pauli matrices (in spinor space) as 
\begin{eqnarray}
\gamma^{0} &=& \begin{pmatrix}
\sigma_{3} & 0\\
0 & - \sigma_{3}
\end{pmatrix},\nonumber\\
\gamma^{1} &=& \begin{pmatrix}
i\sigma_{1} & 0\\
0 & -i\sigma_{1}
\end{pmatrix},\nonumber\\
\gamma^{2} &=& \begin{pmatrix}
i\sigma_{2} & 0\\
0 & - i\sigma_{2}
\end{pmatrix},\label{gammamatrix}
\end{eqnarray} 
and we have identified the effective mass matrix for this constant gauge potential background as $M = m + g \slashed{A}$. For later use, we note here that in this reducible spinor representation, it is possible to define two matrices that anti-commute with the three gamma matrices in \eqref{gammamatrix}, namely,
\begin{eqnarray}
\gamma^{3} &=& i \begin{pmatrix}
0 & 1\\
1 & 0
\end{pmatrix},\nonumber\\\gamma_{5} &=& i\gamma^{0}\gamma^{1}\gamma^{2}\gamma^{3} = i\begin{pmatrix}
0 & 1\\
-1 & 0
\end{pmatrix},\label{gamma5}
\end{eqnarray}
with $\gamma_{5}^{\dagger} = \gamma_{5}, \gamma_{5}^{2} = 1$.
Note that the fermions belong to the fundamental representation of the color $SU(2)$ group and, therefore, carry both the color index $i=1,2$ as well as the spinor index $\alpha=1,2,3,4$. Therefore, the mass matrix $M$ is a $8\times 8$ matrix in the product space and is not Hermitian, $M \neq M^{\dagger}$, since the Dirac matrices $\gamma^{\mu}$ are not Hermitian.  As a result, the eigenvalues of the mass matrix become complex. The eight eigenvalues are easily determined as two complex conjugate pairs 
\begin{align}
m_1 &=m_5= m -  \frac{g}{2} \sqrt{\epsilon_1} - \frac{i}{2} g \sqrt{\epsilon_2} = m_{2}^{*}=m_{6}^{*},  \nonumber
\\ 
m_3 &=m_7= m +  \frac{g}{2} \sqrt{\epsilon_1} - \frac{i}{2} g \sqrt{\epsilon_2} = m_{4}^{*}=m_{8}^{*}, \label{5}
\end{align}
It is worth noting that the chromoelectric field $\epsilon$ is proportional to the real and imaginary parts of the mass eigenvalues, which means that if we had real eigenvalues for the mass matrix, then we would not have a chromoelectric field and, hence, we would not have a decay probability. Since the mass matrix is not self-adjoint, neither is the Hamiltonian of the system, and this is crucial in order to have a nonvanishing decay probability. The decay probability is related to the imaginary part of the effective action and, when fermions are integrated, at one loop level the effective action has the form
\begin{equation}
\Gamma_{\rm eff} =  {\rm Tr}\, \ln \left(i \slashed{D} - m\right).\label{5a}
\end{equation}
We can use the fact that $\gamma_{5}^{2} =1$ and $\gamma_{5}$ anti-commutes with the three gamma matrices (together with the cyclicity of trace) to obtain
\begin{multline}
\Gamma_{\rm eff} = {\rm Tr}\, \ln \left(i \slashed{D} - m\right) =  {\rm Tr}\,\ln \left(i \slashed{D} +m\right)\\ =  \frac{1}{2} {\rm Tr}\, \ln \left(-\slashed{D}^{2} - m^{2}\right).\label{effact}
\end{multline}
Since covariant derivatives $D_{\mu}$ do not commute, we have 
\bb
\slashed{D}^{2} = D_{\mu}D^{\mu} + \frac{g}{2} \sigma^{\mu\nu} F_{\mu\nu}, \label{pauli}
\ee
where we have identified 
\begin{equation}
\sigma^{\mu\nu} = \frac{i}{2} [\gamma^{\mu}, \gamma^{\nu}],
\end{equation}
and the field strength $F_{\mu\nu}$ is defined in \eqref{fs}.

Using these results as well as a proper time representation for the logarithm in \eqref{effact}, the calculation of the pair decay probability by unit of volume and time  can be done as follows; if $w (x)$ is the creation pair probability then 

\begin{multline}
w(x)  =\mbox{Re}\,\, {\mbox tr}_{a \alpha }\left[ \int_0^\infty \frac{ds}{s}\,\langle x, M \mid e^{-is m^2}\right.\\\times\left(\exp\left\{is\left[(p-gA)^2-\frac{g}{2}\sigma_{\mu\nu}F^{\mu\nu}\right]\right\}\right.\\ \left.\left.-e^{-isp^2}\right)\mid x,M\rangle\right]. \label{6}
\end{multline}
 
where the trace  runs over color and spinor indices and $|x,M>$ are  eigenstates of the mass operator ${\hat M}$. A sum over the mass eigenvalues is implicit. 

The last term in RHS corresponds to the subtraction of the  $A=0$ part of the propagator which is equivalent to remove the zero mode.  

%%%

In order to compute the trace in the previous equation one should expand the exponential in a Taylor series. The first exponential has the form
\bb{\rm e}^{is\left[(p-gA)^2-\frac{g}{2}\sigma_{\mu\nu}F^{\mu\nu}\right]}={\rm e}^{\alpha_0+\alpha_1\lambda_1+\alpha_2\lambda_2+\alpha_3\lambda_3},\label{15} \ee
where $\lambda_i$ are the Pauli matrices in color space and $\alpha_i$ are complex matrices in Lorentz space that commute with each other. Recalling that the square of any Pauli matrix is the identity and that the trace of any Pauli matrix vanishes, it is easy to compute the trace on the Taylor series. As a result we are left with another Taylor series
\begin{multline}\tr_{\alpha a}{\rm e}^{\alpha_0+\alpha_1\lambda_1+\alpha_2\lambda_2+\alpha_3\lambda_3}\\=8\left[1+\alpha_0+\frac{1}{2}(\alpha_0^2+\alpha_1^2+\alpha_2^2+\alpha_3^2)\right.\\\left.+\frac{1}{3!}(\alpha_0^3+3\alpha_0\alpha_1^2+3\alpha_0\alpha_2^2+3\alpha_0\alpha_3^2)+\cdots\right].\end{multline}
This new Taylor series can be reconstructed and it can be checked that, finally
\bb\tr_{\alpha a}{\rm e}^{\alpha_0+\alpha_1\lambda_1+\alpha_2\lambda_2+\alpha_3\lambda_3}=8{\rm e}^{\alpha_0}\cosh\sqrt{\alpha_1^2+\alpha_2^2+\alpha_3^2}.\label{tr2}\ee
Taking the definitions of $\alpha_i$ from (\ref{15}) we have

%%%

\begin{multline}\tr_{\alpha a}\left\{e^{is\left[(p+gA)^2-\frac{g}{2}\sigma_{\mu\nu}F^{\mu\nu}\right]}\right\}=8{\rm e}^{is\left[p^2+g^2A^2\right]}\\\times\cosh\sqrt{-s^2g^2\epsilon_1p_0^2-s^2g^2\epsilon_2p_1^2+\left(\frac{sg\epsilon}{2}\right)^2}.\label{tracess}\end{multline}

This result can be confirmed using a calculation software such as Mathematica. Going back to the probability density, we have
\begin{multline}\omega(x)=8\mbox{Re}\int_0^\infty\frac{ds}{s}\langle x, M\mid{\rm e}^{is(p^2-m^2+g^2A^2)}\\\times\cosh\sqrt{-s^2g^2\epsilon_1p_0^2-s^2g^2\epsilon_2p_1^2+\left(\frac{sg\epsilon}{2}\right)^2}\\-{\rm e}^{is(p^2-m^2)}\mid x, M\rangle\end{multline}

Defining $\tilde{m}^2\equiv m^2-g^2A^2$ and taking the real part
\begin{multline}\omega(x)=8\int_0^\infty\frac{ds}{s}\langle x,M\mid\cos\left(s(p^2-\tilde{m}^2)\right)\\\times\cosh\sqrt{-s^2g^2\epsilon_1p_0^2-s^2g^2\epsilon_2p_1^2+\left(\frac{sg\epsilon}{2}\right)^2}\\-\cos\left(s(p^2-m^2)\right)\mid x,M\rangle\end{multline}

We can insert completeness relations between $p$'s and compute the momentum integrals to write, for the second term of $\omega(x)$
\begin{multline}
8\int_0^\infty\frac{ds}{s}\langle x,M\mid\cos\left(s(p^2-m^2)\right)\mid x,M\rangle\\=\frac{1}{(2\pi)^{3/2}}\int_0^\infty\frac{ds}{s^{5/2}}\left[{\rm e}^{ism^2}+{\rm e}^{-ism^2}\right.\\\left.+i{\rm e}^{ism^2}-i{\rm e}^{-ism^2}\right].\label{567}\end{multline}

The calculation of the integrals appearing in (\ref{567}) can be performed observing that they can be regularized by considering the following identity 
\begin{equation}
\lim_{\alpha \to 0} \left[  \int_0^\infty ds \, s^{\gamma -1} \, e^{-\frac{\alpha}{s} - \beta \, s}\right] = \frac{2^{1-\gamma}}{\beta^\gamma},\label{reg}  
\end{equation}
where $\beta$ now is $\pm im^2$ and $\gamma=-3/2$. With this, the integral in (\ref{567}) will vanish and the probability density becomes
\begin{multline}\omega(x)=8\int_0^\infty\frac{ds}{s}\int\frac{d^3p}{(2\pi)^3}\cos\left(s(p^2-\tilde{m}^2)\right)\\\times\cosh\sqrt{-s^2g^2\epsilon_1p_0^2-s^2g^2\epsilon_2p_1^2+\left(\frac{sg\epsilon}{2}\right)^2}. \label{prob1}\end{multline}

The integral in (\ref{prob1}) cannot be computed exactly. Rather we will look for an upper bound for the probability density. Such an upper bound can be obtained by making the replacement
\begin{multline}\cosh\sqrt{-s^2G_1^2p_0^2-s^2G^2p_1^2+\left(\frac{sg\epsilon}{2}\right)^2}\\\rightarrow\cosh\left(\frac{sg\epsilon}{2}\right)\label{cotas}.\end{multline}
The momentum integrals can then be computed and the regularization shown in (\ref{reg}) can be performed. We define

\begin{eqnarray}
R&=&\sqrt{\left(\frac{g\epsilon}{2}\right)^2+(\tilde{m}^2)^2}\\
\theta&=&\arctan\left(\frac{\tilde{m}^2}{g\epsilon/2}\right),
\end{eqnarray}

and obtain for the upper bound of the probability density
\begin{multline}\omega<\frac{R^{3/2}}{4(2\pi)^{3/2}}\left\{(1+i)\left[{\rm e}^{\frac{3}{2}i(\theta\mp\pi)}+{\rm e}^{-\frac{3}{2}i\theta}\right]\right.\\\left.+(1-i)\left[{\rm e}^{\frac{3}{2}i(-\theta\pm\pi)}+{\rm e}^{\frac{3}{2}i\theta}\right]\right\},\label{up}\end{multline}
where the upper sign in $(\mp\pi,\pm\pi)$ is for $\tilde{m}^2\geq0$ and the lower sign is for $\tilde{m}^2<0$. It easily verified that for $\tilde{m}^2\geq0$ the previous expression vanishes. So, in order to have a non vanishing probability density $\tilde{m}^2<0$, i.e.
\bb m^2<g^2A^2=\frac{g^2}{4}(\epsilon_1-\epsilon_2).\label{cond}\ee
With this, the upper bound for our probability density is
\bb\omega<\left(\frac{R}{2\pi}\right)^{3/2}\left(\cos\frac{3}{2}\theta+\sin\frac{3}{2}\theta\right).\ee
A similar process can be carried out and a lower bound for the probability density can be found through the replacement
\begin{multline}\cosh\sqrt{-s^2G_1^2p_0^2-s^2G^2p_1^2+\left(\frac{sg\epsilon}{2}\right)^2}\\\rightarrow\cos\left(sG_1p_0+sGp_1\right).\label{cotai}\end{multline}
The computation of the lower bound can then be performed and it turns out to be null, so we have
\bb0<\omega<\left(\frac{R}{2\pi}\right)^{3/2}\left(\cos\frac{3}{2}\theta+\sin\frac{3}{2}\theta\right).\ee
It is important to note that in making the replacements in (\ref{cotas}) and (\ref{cotai}) we have asumed that the probability density is positive definite. This restricts the region where our bounds are valid and the possible values of $\theta$. Taking this into account, as well as the condition in (\ref{cond}) we can find the region of allowed values for $\theta$
\begin{multline}
-\frac{(8n+1)\pi}{6}<\theta<-\frac{(8n-3)\pi}{6}\\\mbox{for}\quad n=1,2,3,\ldots\end{multline}
or
\begin{equation}
-\frac{\pi}{6}<\theta<0. 
\end{equation}
With this, we can plot the allowed region whithin which the probability density has a finite, nonvanishing value. This is shown in Fig. 1.

\begin{figure}
\begin{center}
\includegraphics[scale=0.25]{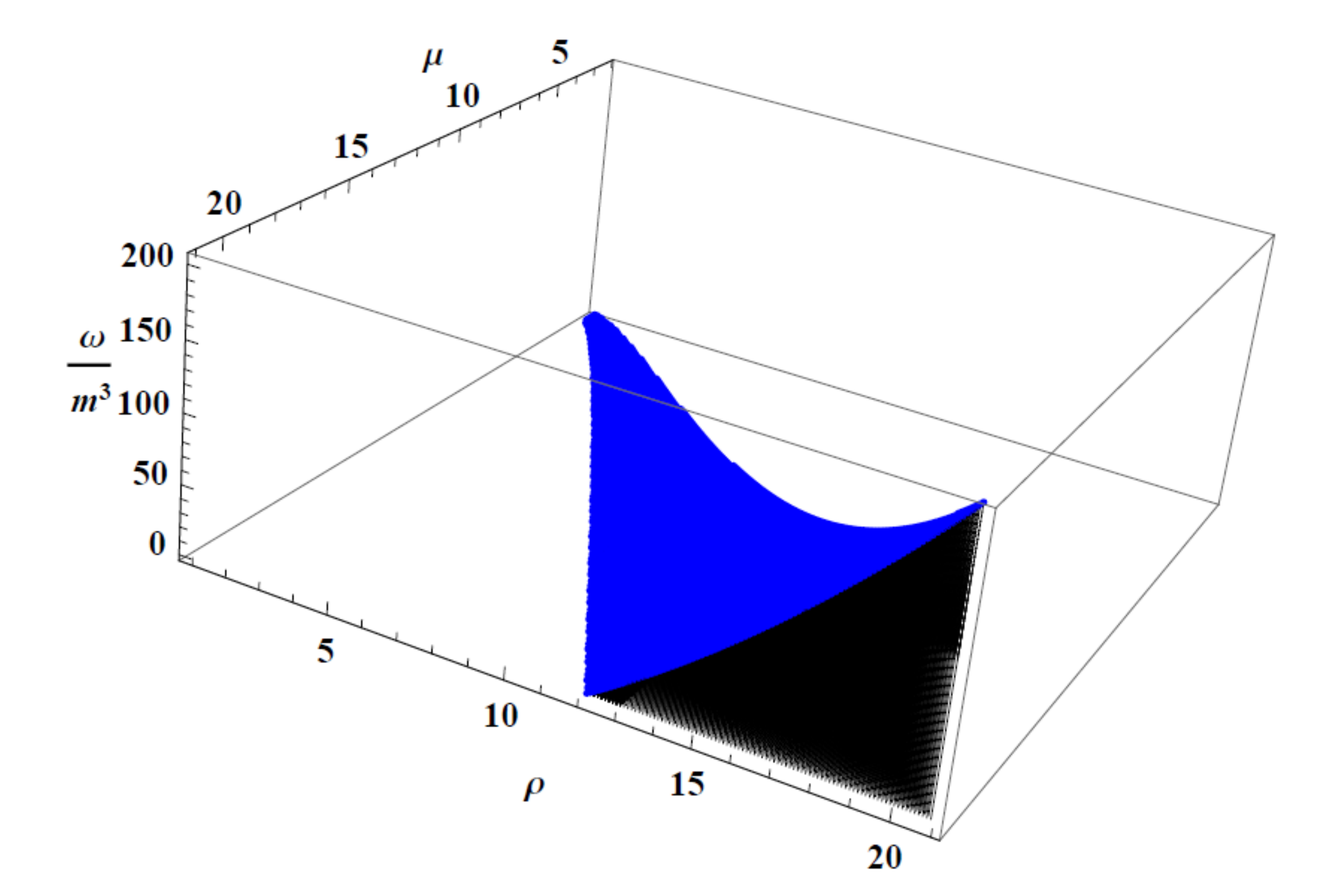}
\end{center}
\caption{Probability density for pair emission in units of $m^3$. The blue surface represents the upper bound. The black region is where the exact value of the probability can be. Dimensionless variables $\mu=\frac{g\sqrt{\epsilon_1}}{m}$ and $\rho=\frac{g\sqrt{\epsilon_2}}{m}$ have been defined.}
\end{figure}

\section{Color SU(3).}

This same problem can be solved within the color $SU(3)$ group. The corresponding potential is the same as in $SU(2)$ with $A_\mu^a\neq 0$ only for $a=1,2$. Notice that this gauge fixing condition implies that actually this is not a full $SU(3)$ calculation, which would imply the appearance of the eight generators. Our calculation corresponds only to the projection of $SU(3)$ into $SU(2)$. The computation of the trace in (\ref{tracess}) is now different

\begin{multline}\tr_{\alpha a}\left\{e^{is\left[(p+gA)^2-\frac{g}{2}\sigma_{\mu\nu}F^{\mu\nu}\right]}\right\}=4{\rm e}^{is\left[p^2+g^2A^2\right]}\left(1+\right.\\\left.2\cosh\sqrt{-s^2g^2\epsilon_1p_0^2-s^2g^2\epsilon_2p_1^2+\left(\frac{sg\epsilon}{2}\right)^2}\right).\end{multline}

If we compare this with (\ref{tracess}) we can see that a new term has been added to our result coming from the first term in the RHS. However, because of the same argument following (\ref{567}) this term will have a vanishing contribution to the probability density. Taking this into account we can say that the probability density in $SU(3)$ color group yields the same result we have already shown.

\section{3+1 dimensions.}

So far we have restricted our analysis to the $2+1$ dimensional case. In this section we will briefly discuss the $3+1$ dimensional case. In such a case we have an additional phase space integral and then Eq. (\ref{567}) is changed like
\begin{multline}
8\int_0^\infty\frac{ds}{s}\langle x,M\mid\cos\left(s(p^2-m^2)\right)\mid x,M\rangle\\=\frac{-i}{(2\pi)^{2}}\int_0^\infty\frac{ds}{s^{3}}\left[{\rm e}^{ism^2}-{\rm e}^{-ism^2}\right].\end{multline}
The extra dimension has increased the exponent of $s$ from $5/2$ to $3$, meaning that the regularization shown in (\ref{reg}) should now be performed taking $\gamma=-2$. This is key since now, Eq. (\ref{up}) will look like
\begin{equation}\omega<\frac{R^{2}}{2(2\pi)^{2}}\left\{{\rm e}^{2i(\theta\mp\pi)}-{\rm e}^{2i\theta}\right\},\end{equation}
which vanishes. This result coincides with that of reference \cite{brown}. It is worth noting that the reason for this null decay probability lies in the fact that we have enlarged our phase space. It is the extra dimension in $s$ what makes the integrals vanish.

\section{Scalar Fields.}

Computations of vacuum instability are usually easier to carry out when one is dealing with scalar fields. Given that we could not obtain an exact expression for the probability density for pair emission when working with fermionic fields, one might ask wether or not this is possible when working with colorful bosonic fields. In such a case, the probability density for pair emission is
\begin{multline}w(x)  =\mbox{Re}\,\, {\mbox tr}_{a}\left[ \int_0^\infty \frac{ds}{s}\,\langle x, M \mid e^{-is m^2}\right.\\\times\left(\exp\left\{is\left[(p-gA)^2\right]\right\}\right.\\ \left.\left.-e^{-isp^2}\right)\mid x,M\rangle\right], \label{6111}\end{multline}
where the trace now runs only over color indices. The trace can be computed in the same manner as before by taking $\alpha_3=0$ in Eq. (\ref{tr2})
\begin{multline}\tr_{a}\left\{e^{is\left[(p+gA)^2\right]}\right\}=2{\rm e}^{is\left[p^2+g^2A^2\right]}\\\times\cosh\sqrt{-s^2g^2\epsilon_1p_0^2-s^2g^2\epsilon_2p_1^2}.\label{tracess111}\end{multline}
Then, after inserting copleteness relations between $p'$s we get
\begin{multline}\omega(x)=8\int_0^\infty\frac{ds}{s}\int\frac{d^3p}{(2\pi)^3}\cos\left(s(p^2-\tilde{m}^2)\right)\\\times\cos\sqrt{s^2G_1^2p_0^2+s^2G^2p_1^2}. \label{prob111}\end{multline}
Once again, this integral cannot be computed exactly. As in the previous sections we will then find an upper and lower bound for it. We can find an upper bound for the probability through the replacement
\begin{equation}\cos\sqrt{s^2G_1^2p_0^2+s^2G^2p_1^2}\rightarrow 1\label{cotass111},\end{equation}
and so we can write
\begin{equation}\omega(x)< 8\int_0^\infty\frac{ds}{s}\int\frac{d^3p}{(2\pi)^3}\cos\left(s(p^2-\tilde{m}^2)\right)\end{equation}
By the same argument following Eq. (\ref{567}), it is easy to show that this last integral vanishes. This means then that the probability density itself vanishes and hence, there is no vacuum instability. In this sense, even though the computation is not much easier when dealing with scalar fields, we are able to conclude that there is no vacuum instability.\\

The same analysis here presented can be carried out in $3+1$ dimensions. It is trivial to show that, in the same way the probability vanished when dealing with fermions in $3+1$ dimensions, it will also vanish when dealing with scalars.\\

\section{Conclusions.}

We analyzed the problem of pair emission from the vacuum in the presence of a chromoelectric field in $2+1$ dimensions. The probability density cannot be computed exactly, but we find an upper and lower bound for it. We also find that there is an allowed region in the parameter space, where the probability will not vanish, shown in Fig. 1. We show that the probability density however, vanishes in $3+1$ dimensions. We also find that for scalar fields, the probability density vanishes both in $2+1$ and $3+1$ dimensions.

\section{Acknowledgments} 

The authors would like to thank J. Gamboa and A. Das for important contributions at the beginning of this work. Also, we would like to acknowledge helpful discussion with Ben Koch, F. M\'endez and A. Raya. Finally, we want to thank S. Gavrilov for helpful and interesting remarks to our work. This  work was supported by FONDECYT-Chile under grants 1130056 and 1120770, and by CONICYT under grants 21110295 (RZ) and 21110577 (FM).

\bibliography{Vacbib}{}
\bibliographystyle{apsrev}

%%%%%%%
\begin{comment}

\end{comment}
\end{document}